\documentclass[twocolumn,twoside,slac_two]{revtex4}
\usepackage{graphicx}
\usepackage{fancyhdr}
\begin{document}

\title{The Cosmological Evolution of Blazars and the Extragalactic Gamma-Ray Background in the Fermi Era}

%

\author{Yoshiyuki Inoue, Tomonori Totani}
\affiliation{Department of Astronomy, Kyoto University, Kitashirakawa, Sakyo-ku, Kyoto 606-8502, Japan}
\author{Susumu Inoue}
\affiliation{Department of Physics, Kyoto University, Kitashirakawa, Sakyo-ku, Kyoto 606-8502, Japan}
\author{Masakazu A. R. Kobayashi}
\affiliation{Optical and Infrared Astronomy Division, National Astronomical Observatory of Japan, Mitaka, Tokyo 181-8588, Japan}
\author{Jun Kataoka}
\affiliation{Waseda University, 1-104 Totsukamachi, Shinjuku-ku, Tokyo 169-8050, Japan}
\author{Rie Sato}
\affiliation{Department of High Energy Astrophysics, Institute of Space and Astronautical Science (ISAS), Japan Aerospace Exploration Agency (JAXA), 3-1-1 Yoshinodai, Sagamihara, 229-8510, Japan}

\begin{abstract}
  The latest determination of the extragalactic gamma-ray background
  (EGRB) radiation by {\it Fermi} is compared with the theoretical
  prediction of the blazar component by Inoue \& Totani (2009;
  hereafter IT09). The {\it Fermi} EGRB spectrum is in excellent
  agreement with IT09, indicating that blazars are the dominant
  component of the EGRB, and contributions from any other sources (e.g.,
  dark matter annihilations) are minor. It also indicates that the
  blazar SED (spectral energy distribution) sequence taken into
  account in IT09 is a valid description of mean blazar SEDs.
  The possible contribution of MeV blazars to the EGRB in the MeV band is also discussed. 
  In five total years of observations, we predict that {\it Fermi} will detect
  $\sim$1200 blazars all sky down to the corresponding sensitivity limit.
  We also address the detectability of the highest-redshift blazars.
  Updating our model with regard to high-redshift evolution
  based on SDSS quasar data,
  we show that {\it Fermi} may find some blazars up to $z\sim6$ during the
  five-year survey. Such blazars could provide a new probe
  of early star and galaxy formation through GeV spectral attenuation
  signatures induced by high-redshift UV background radiation.
\end{abstract}

\maketitle

\thispagestyle{fancy}


\section{Introduction}
\label{intro}

The origin of the extragalactic diffuse gamma-ray background (EGRB)
is a long-standing puzzle. First discovered by the
\textit{SAS 2} satellite \cite{FST78,TF82},
its existence was subsequently confirmed up to $\sim$ 50 GeV by EGRET (Energetic
Gamma-Ray Experiment Telescope) on board the Compton Gamma Ray
Observatory. EGRET measured a flux of about $1 \times$ 10$^{-5}$ photons
cm$^{-2}$ s$^{-1}$ sr$^{-1}$ above 100 MeV and an approximately power-law spectrum with photon index of $\sim -2$ over the wide
range of 30 MeV -- 50 GeV \cite{sre98,SMR04a}. Very recently,
Ref. \cite{ack09} reported {\it Fermi} observations of the EGRB spectrum up to 100 GeV,
which connects smoothly to the EGRET results below $\sim$ 200 MeV \cite{SMR04a}.
However, above this energy, the spectrum is discrepant from EGRET,
being a single power-law with index $\sim -2.4$,
and showing no signs of a GeV excess.

Although different types of gamma-ray sources (e.g., clusters of galaxies or
dark matter annihilation) have been proposed to be significant
contributors to the EGRB [see, e.g., Ref. \cite{NT06} and references
therein], active galactic nuclei (AGNs) of the blazar class are considered the primary candidates, since almost all of the extragalactic sources detected by EGRET were
blazars. The blazar contribution to the EGRB has been estimated
by numerous authors,
who have reached different conclusions using different approaches,
ranging from 20\% to 100 \% of the observed flux
\cite{PGF93,SSM93,SS94,chi95,SS96,CM98,MP00,NT06,gio06,D07,PV08,KM08,BSM09}.

In most past studies, the spectral energy distributions
(SEDs) of the blazars were assumed to be power laws for all objects. In such
models, it is obvious that the predicted EGRB spectrum is mainly determined
by the assumed power-law indices. However, multi-wavelength observational studies
indicate the existence of a non-trivial trend among blazar SEDs:
the energies of the two characteristic spectral peaks in blazar SEDs
(each likely reflecting the synchrotron and inverse Compton emission)
systematically decrease as the bolometric luminosity increases
\cite{fos97,fos98,kub98,ghi98,don01,GT08,mar08}. This is often referred to as the
blazar SED sequence.  Although its validity is currently still a matter of debate (e.g.,
Ref. \cite{pad07,GT08,mar08}), one can make a non-trivial prediction
of the EGRB spectrum if this blazar sequence is assumed.

Here we compare the latest {\it Fermi} EGRB results
with our model predictions, previously published in
Ref. \cite{IT09} (IT09) before the former was announced.
IT09 calculated the EGRB from blazars by constructing a
blazar gamma-ray luminosity function (GLF) model that is consistent
with the flux and redshift distributions of the EGRET blazars, accounting for
the blazar sequence as well as the luminosity-dependent density evolution
(LDDE) scheme that describes well the evolution of the X-ray luminosity
function (XLF) of AGNs. By introducing the blazar SED sequence, we
were able to make a reasonable and non-trivial prediction of the EGRB
spectrum that can be compared with observations including the new {\it Fermi} data.

Recently, Ref. \cite{ITU08} has shown that the EGRB in the MeV band can
be naturally explained by normal (i.e., non-blazar) AGNs that compose
the cosmic X-ray background. If so, they should also contribute to
the EGRB at $\lesssim$ 1 GeV. We will investigate what fraction of
the observed EGRB can be explained by the sum of the blazar and non-blazar AGN components.
The contribution of blazars to the EGRB in the MeV band is also discussed
and compared with a recent study on the possible relevance of the
so-called ``MeV blazars'' \cite{aje09}.

We will then make quantitative predictions for the {\it Fermi}
Gamma-ray Space Telescope \cite{atw09} for its five-year survey. The
first {\it Fermi} catalog for bright gamma-ray sources including AGNs
has recently been released \cite{abd09a,abd09b}. 
The number of currently detected blazars is about 100, larger than that of EGRET
by a modest factor of about 2. We limit the discussion
of the blazar data to only the EGRET sample in this work,
to be considered as a theoretical prediction from the pre-{\it Fermi} era.
In the future, a much larger number of blazars should be detected with {\it Fermi},
which can be compared further with our predictions.

We also discuss the highest-redshift blazars detectable by {\it Fermi},
by updating the IT09 GLF model with respect to high-redshift evolution.
We briefly mention
the possibility of utilizing such blazars to obtain information
on the high-redshift extragalactic background light and early galaxy formation
through spectral absorption features.

Throughout this paper, we adopt the standard cosmological parameters
of ($h,\Omega_M,\Omega_\Lambda$)=(0.7,0.3,0.7).

\section{The Blazar SED Sequence}
\label{sequence}

\begin{figure}
\centering
\includegraphics[width=70mm]{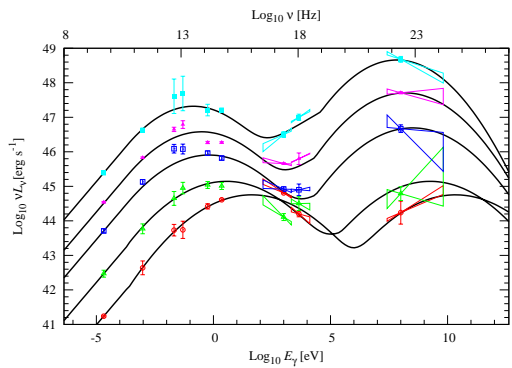}
\caption{The blazar SED sequence. The data points are the average SED of the blazars studied by Ref. \cite{fos98,don01}. The solid curves are the empirical SED sequence models constructed and used in this paper. The model curves corresponds to the bolometric luminosities of $\log_{10} (P/{\rm erg \ s^{-1}}) = $ 49.50, 48.64, 47.67, 46.37, and 45.99 (from top to bottom).}
\label{fig.sed}\end{figure}

Ref. \cite{fos97,fos98,don01} constructed an empirical blazar SED model to describe the SED sequence, based on fittings to observed SEDs from radio to $\gamma$-ray bands. These models are comprised of the two components (synchrotron and IC), and each of the two is described by a linear curve at low photon energies and a parabolic curve at high energies.

We construct our own SED sequence model mainly based on the SED model \cite{don01}, because there is a mathematical discontinuity in the original model of Ref. \cite{don01}. Our own SED sequence formula is described in the Appendix of Ref. \cite{IT09} in detail. In Fig.\ref{fig.sed}, we show this empirical blazar SED sequence model in comparison with the observed SED data \cite{fos98,don01}.

\section{The Model of Gamma-ray Luminosity Function of Blazars}
\label{glf_model}
The cosmological evolution of AGN XLF has been investigated intensively \cite{Ued+03,HMS05,Saz+07,GCH07}. These studies revealed that AGN XLF is well described by the LDDE model, in which peak redshift of density evolution increases with AGN luminosity. Here we construct blazar GLF models based on the two XLFs derived by Ref. \cite{Ued+03} (hereafter U03; in hard X-ray band) and Ref. \cite{HMS05} (hereafter H05; in soft X-ray band).  The use of LDDE in blazar GLF has been supported from the EGRET blazar data \cite{NT06}.

We simply assume that the bolometric luminosity of radiation from jet, $P$, is proportional to disk X-ray luminosity, $L_X$.  We relate these two by the parameter $q$, as $P=10^qL_X$. Here, we define the disk luminosity $L_X$ to be that in the rest-frame 2--10 and 0.5--2 keV bands for the U03 XLF and the H05 XLF
, respectively. Thus, the luminosity at rest 100 MeV, $L_\gamma$, and $L_X$ have been related through $P$.

The blazar GLF $\rho_\gamma$ is then obtained from the AGN XLF, $\rho_X$, as $\rho_\gamma(L_\gamma, z) = \kappa\frac{dL_X}{dL_\gamma}\rho_X(L_X, z)$, where $\rho_\gamma$ and $\rho_X$ are the comoving number densities per unit gamma-ray and X-ray luminosity, respectively.  The parameter
$\kappa$ is a normalization factor, representing the fraction of AGNs observed as blazars. See the \S 3 in Ref. \cite{IT09} for a detail.

\begin{figure*}
  \begin{center}
    \begin{tabular}{cc}
      \resizebox{70mm}{!}{\includegraphics{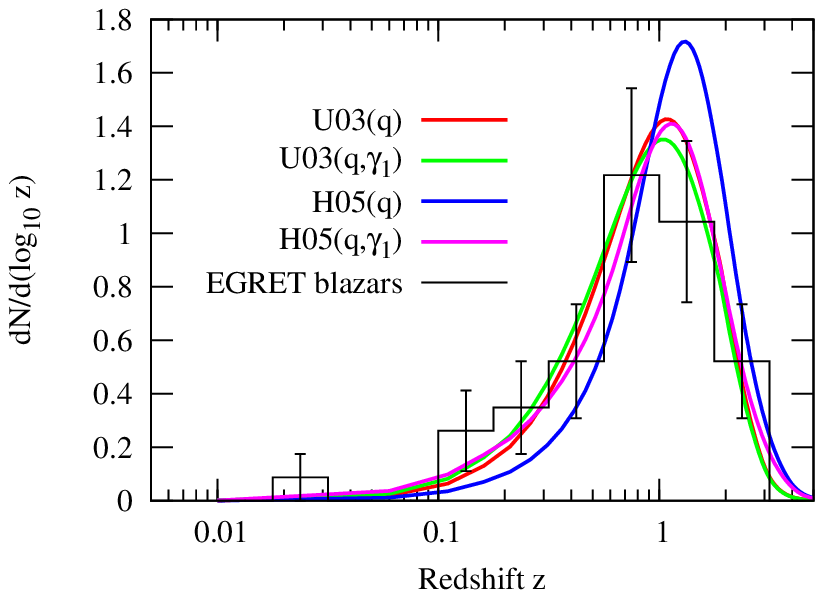}} &
      \resizebox{70mm}{!}{\includegraphics{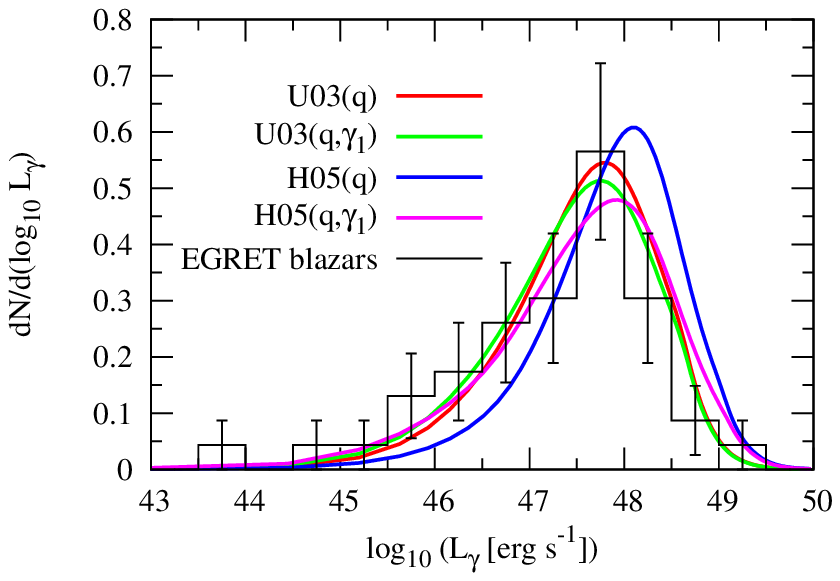}} \\
    \end{tabular}
    \caption{Redshift and gamma-ray luminosity ($\nu L_\nu$ at 100 MeV) distributions of EGRET blazars. The histogram is the binned EGRET data, with one sigma Poisson errors.  The four model curves are the best-fits for the GLF models of U03($q$), H05($q$), U($q,\gamma_1$), and H05($q,\gamma_1$), as indicated in the figure.  Bottom panels: the same as the top panels, but for cumulative distributions.}
    \label{EGRET_dist}
  \end{center}
\end{figure*}

We use the maximum likelihood method to search for the best-fit model parameters of the blazar GLF to the distributions of the observed quantities of the EGRET blazars (gamma-ray flux and redshift). See Ref. \cite{IT09} for a detail. We take $q$ and the faint-end slope index of XLF $\gamma_1$ in addition to $q$ as free parameters . These are hereafter called as U03($q$), H05($q$), U03($q$, $\gamma_1$) and H05($q$, $\gamma_1$) fits, respectively.   Figure \ref{EGRET_dist} shows the distributions of redshift and gamma-ray luminosity predicted by the best-fit models, in comparison with the EGRET data.

We performed the Kolmogorov-Smirnov (KS) test to see the goodness of fits for the best-fit results of each of the four fits, and the chance probabilities of getting the observed KS deviation. Since the best KS test value is obtained for the U03($q, \gamma_1$) GLF model, we use this GLF model as the baseline below.

\section{The EGRB Spectrum}
\label{section:egrb}

We calculate the EGRB spectrum by integrating our blazar SED sequence model using the blazar GLF
derived in \S \ref{glf_model}. Since EGRET has already resolved bright gamma-ray sources, we should not include those sources in the EGRB calculation. Therefore, the maximum gamma-ray luminosity in the integration should be $L_{\gamma}(F_{\rm EGRET},z)$, which is the luminosity of the blazar having a flux $F_\gamma = F_{\rm EGRET}$ at a given $z$, where $F_{\rm EGRET}=7 \times 10^{-8} \ \rm photons \ cm^{-2} \ s^{-1}$ is the EGRET sensitivity above 100 MeV.

High energy photons ($\gtrsim$ 20 GeV) from high redshift are absorbed by the interaction with the cosmic infrared background (CIB) radiation \cite{SaSt98,TT02,KBMH04,SMS06}.  We adopt the model of Ref. \cite{TT02} for the CIB optical depth. We also take into account the cascading emission to calculate EGRB spectrum, considering only the first generation of created pairs \cite{KM08}. We found that the amount of energy flux absorbed and reprocessed in intergalactic medium is only a small fraction of the total EGRB energy flux , and hence the model dependence of CIB or the treatment of the cascading component does not have serious effects on our conclusions in this work \cite{IT09}.

In addition to blazars, we take into account non-blazar AGNs as the source of EGRB. Ref. \cite{ITU08} (hereafter ITU08) has shown that non-blazar AGNs are a promising source of EGRB at $\sim$1--10 MeV.  In this scenario, a nonthermal power-law component extends from hard X-ray to $\sim$10 MeV band in AGN spectra, because of nonthermal electrons that is assumed to exist in hot coronae around AGN accretion disks. The existence of such nonthermal electrons is theoretically reasonable, because magnetic reconnection is the promising candidate for the heating source of the hot coronae \cite{LB08} where hard X-ray photons are emitted. Magnetic reconnections should produce nonthermal particles, as is well known in e.g., solar flares or the earth magnetosphere.  Although several sources have been proposed as the origin of the MeV background, this model gives the most natural explanation for the observed MeV background spectrum that is a simple power-law {\it smoothly} connected to CXB.

Theoretically, there is no particular reason to expect a cut-off of the nonthermal emission around 10 MeV, and it is well possible that this emission extends beyond 10 MeV with the same power index.  Then, this component could make some contribution to GeV EGRB.  The EGRB spectra from the non-blazar AGNs are calculated based on the model of ITU08.  The model parameters of ITU08 have been determined to explain the EGRB in the MeV band.  There is some discrepancy between the MeV EGRB data of SMM \cite{SMM} and COMPTEL \cite{COMPTEL}, and the reason for this is not clear.  Here we use two model parameter sets of $(\Gamma, \gamma_{\rm tr})=(3.5,4.4)$ and $(3.8,4.4)$ in ITU08, where $\Gamma$ is the power-law index of nonthermal electron energy distribution and $\gamma_{\rm tr}$ is the transition electron Lorentz factor above which the nonthermal component becomes dominant.  These two parameter sets are chosen so that they fit to the COMPTEL and SMM data, respectively.

\subsection{On the Origin of the EGRB: Comparison with the {\it Fermi} EGRB Spectrum}
\label{subsec:origin_egrb}

\begin{figure}
  \begin{center}
   \includegraphics[width=70mm]{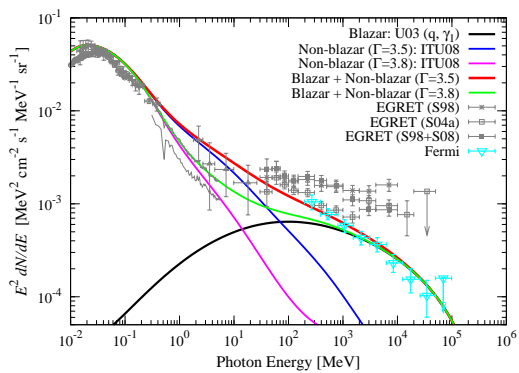}
    \caption{The EGRB spectrum from non-blazar AGNs and blazars for the GLF model of U03($q,\gamma_1$).
    The model curves for the blazar component (absorbed+cascade), non-blazar AGN component,
    and the sum of the two are shown. Note that two models are plotted for the non-blazar component
    with different values of $\Gamma$.  The observed data of HEAO-1 \cite{HEAO1}, {\it Swift}/BAT \cite{BAT},
    SMM \cite{SMM}, COMPTEL \cite{COMPTEL} and EGRET \cite{sre98,SMR04a} are shown.
    We also plot an alternative version of the EGRET data denoted as ``S98+S08''
    which is the original EGRET results \cite{sre98} modified with the correction factors of Ref. \cite{SHK08}.
    The {\it Fermi} EGRB data is also shown \cite{ack09}. }
    \label{fig.egrb}
  \end{center}
\end{figure}

Fig. \ref{fig.egrb} compares our model predictions with the observed EGRB data,
including the very recent {\it Fermi} results \cite{ack09}
\footnote{Note that the {\it Fermi} EGRB data was announced after the submission of Ref. \cite{IT09}.}.
Since the flux thresholds of EGRET and {\it Fermi} are different,
we must be careful when directly comparing with the {\it Fermi} data.
However, when the sensitivity threshold of {\it Fermi} is set to be the same as that of EGRET,
the EGRB flux changes by only $\sim$10 \% \cite{ack09_fermi}.
Therefore, our predicted EGRB spectrum can be considered to be in excellent agreement
with the {\it Fermi} data. This agreement strongly suggests that blazars are the principal sources of
the EGRB above 1 GeV, and the contribution from other sources like dark matter annihilation is minor.
This also implies that use of the blazar sequence is a valid assumption.

The EGRB prediction using the ITU08 non-blazar background model with $\Gamma = 3.5$
also agrees nicely with the observed data from X-rays to 100 GeV.
In this case, the predicted EGRB flux above 100 MeV can account for 80\% of the observed flux, considerably higher than in previous studies \cite{CM98,MP00,NT06}.
It should be noted that the contribution to the EGRB flux above 100 MeV from blazars alone is $\sim$45 \%,
while the remaining $\sim$35 \% comes from non-blazar AGNs.

Based on these results and considerations, we conclude that most, and probably all,
of the EGRB flux can be explained by blazars together with non-blazar AGNs,
whose luminosity functions are consistent with the EGRET blazar data and X-ray AGN surveys.

\subsection{Non-blazar AGNs and Blazars}

As can be seen in Fig. \ref{fig.egrb}, the spectrum of the cosmic MeV background
connects smoothly with that of the cosmic X-ray background
(CXB), while the spectral index of the EGRB changes abruptly around
$\sim$10 MeV. This indicates that the sources of the EGRB below 10
MeV is the same population as that of the CXB (i.e. non-blazar AGNs),
and the sources of the EGRB above 10 MeV is a different population (i.e. blazars).

Recently, based on observations of Swift/BAT-selected blazars at 15--55 keV,
Ref. \cite{aje09} suggested that MeV blazars 
contribute significantly to the MeV background.
However, this conclusion is based on an extrapolation of blazar SEDs
from 55 keV to the MeV band, and some fine tuning may be required in
the MeV blazar SED to reproduce the EGRB spectrum that connects smoothly with the CXB.
Moreover, the spectral break in the EGRB around 10 MeV implies that such
MeV blazars are a distinct population from those contributing to
the EGRB above 10 MeV.  Therefore, it seems more natural that the
dominant contribution to the MeV background is coming from non-blazar AGNs
as proposed by \cite{ITU08}.  Further observational studies are
desired to reveal the true origin of the MeV background.

\section{Predictions for the {\it Fermi} mission}
\label{section:Fermi}

\subsection{Expected Number of Blazars and Non-blazar AGNs}

\begin{figure*}
  \begin{center}
    \begin{tabular}{cc}
      \resizebox{70mm}{!}{\includegraphics{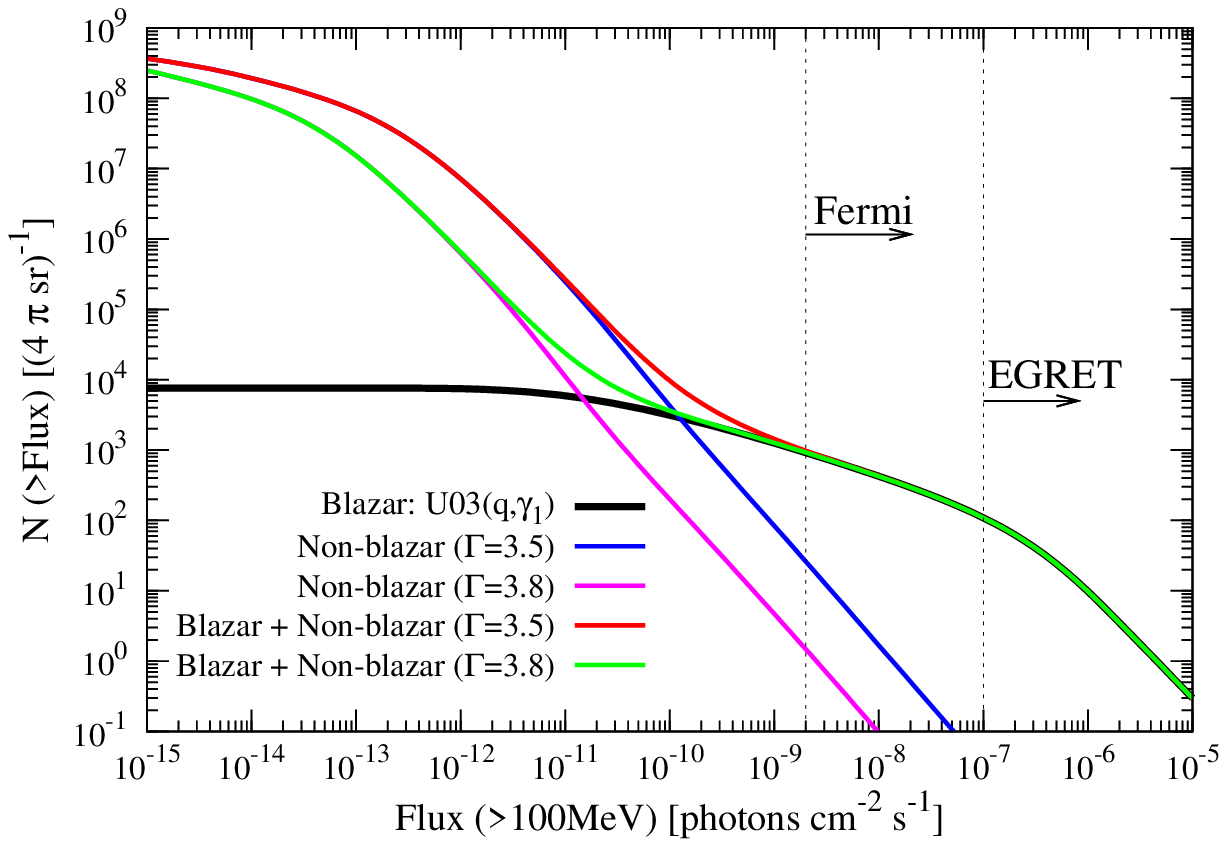}} &
      \resizebox{70mm}{!}{\includegraphics{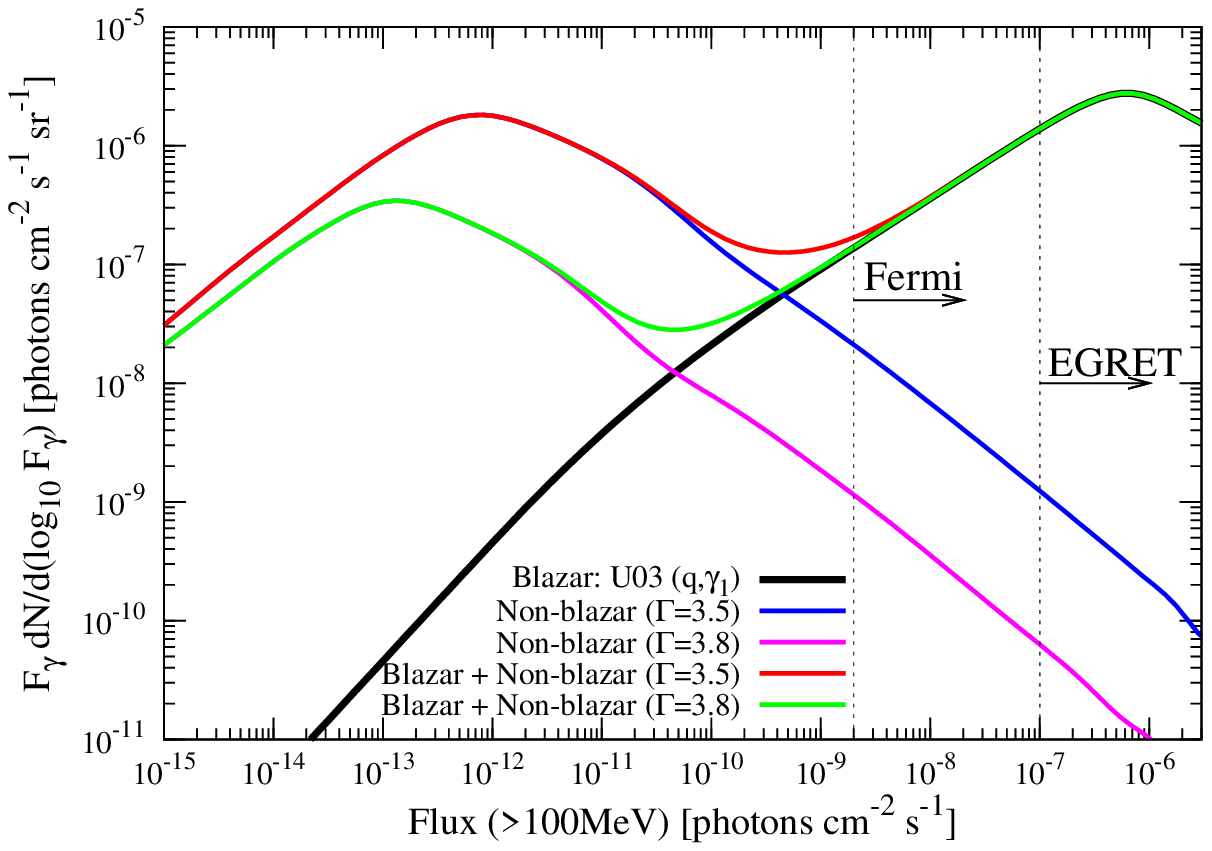}} \\
    \end{tabular}
    \caption{$Left$: the cumulative flux distributions of blazars, non-blazar AGNs, and the total of the two.  The detection limits of EGRET and {\it Fermi} are also shown. $Right$: the same as the left panel, but showing differential flux distribution multiplied by flux $F_\gamma$, to show the contribution to EGRB per logarithmic flux interval. The two models of non-blazar AGNs with different values of $\Gamma$ are shown. The total of blazar and non-blazar counts is also shown.}
  \label{fig.count}
  \end{center}
\end{figure*}

The left panel of Figure \ref{fig.count} shows the cumulative distribution of $>$ 100 MeV photon flux of blazars, non-blazar AGNs, and the total of the two. The GLF predicts that about 750 and 1200 blazars
should be detected by {\it Fermi} where we assumed the {\it Fermi} sensitivity to be $F_{\rm lim}=3\times 10^{-9}$ and $1\times 10^{-9}$ photons cm$^{-2}$ s$^{-1}$ above 100 MeV. These sensitivities correspond to $\sim$1 and $\sim$5 years survey sensitivities.
 
The model of Ref. \cite{SS96}, and the PLE and LDDE models \cite{NT06} predicted $\sim$10000, 5000, and 3000 blazars for the 1-year sensitivity limit, respectively.  It is remarkable that the GLF models in this work predict significantly smaller numbers of blazars than most of previous studies.

About 4--50 non-blazar AGNs would be detected by {\it Fermi} with 5-year survey, by their soft nonthermal emission from nonthermal electrons in coronae of accretion disks, giving an interesting test for our model.  We estimated the $>$ 100 MeV gamma-ray flux from the observed hard X-ray flux of NGC 4151 (the brightest Seyfert galaxy in all sky \cite{Saz+07}) using the ITU08 model \cite{IT09}. Then, NGC 4151 is marginally detectable when $\Gamma = 3.8$, and easily detectable with $\Gamma = 3.5$ with one year survey sensitivity.

The right panel of Figure \ref{fig.count} shows the differential flux distribution of gamma-ray blazars and non-blazar AGNs multiplied by flux, showing the contribution to the EGRB per unit logarithmic flux interval. EGRB from blazars should practically be resolved into discrete sources.  We find that the fraction of EGRB flux that should be resolved is 98\% against  the total blazar EGRB flux in 5-year survey.  On the other hand, EGRB from non-blazar AGNs is hardly resolved into discrete sources by {\it Fermi}.  The predicted fraction of non-blazar EGRB flux resolved by {\it Fermi} is $\sim$0.01 \% with 5-year survey, respectively, against the total non-blazar EGRB flux.  These results are in sharp contrast to those for blazars, although the contributions to the total EGRB by the two populations are comparable at around 100 MeV.

The above results mean that a considerable part of EGRB at around 100 MeV will remain unresolved even with the {\it Fermi} sensitivity, because there is a considerable contribution from non-blazar AGNs to EGRB at $\sim$100 MeV.  However, the contribution from non-blazars should rapidly decrease with increasing photon energy, and almost all of the total EGRB flux at $\gtrsim$ 1 GeV should be resolved into discrete blazars by {\it Fermi}, if there is no significant source contributing to EGRB other than blazars.  It should be noted that these predictions can be tested by {\it Fermi} relatively easily, without follow-up or cross check at other wavebands, once measurements of source counts and EGRB flux have been done by the {\it Fermi} data. Therefore this gives a simple and clear test for our theoretical model \cite{IT09}.

\subsection{High Redshift {\it Fermi} Blazars}

\begin{figure}
  \begin{center}
   \includegraphics[width=70mm]{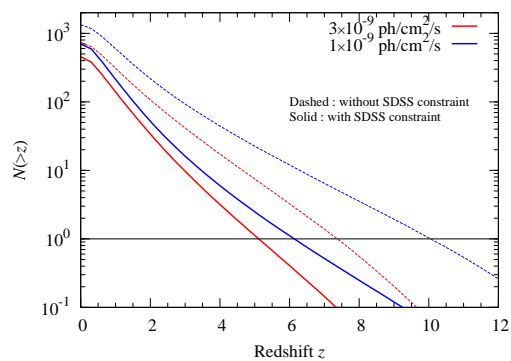}
    \caption{Expected cumulative redshift distribution of blazars detectable by {\it Fermi} above 100 MeV. Red and blue curves correspond
    to the flux sensitivity limits of $F_{\rm lim}$ ($>$100MeV) $=3\times10^{-9}$ and $1\times10^{-9}$
    in units of photons cm$^{-2}$ s$^{-1}$. Solid and dashed curves correspond to whether or not SDSS quasar constraints are taken into account.}
    \label{fig.z_count}
  \end{center}
\end{figure}

Since {\it Fermi} is the most sensitive GeV gamma-ray observatory to date,
it has the potential to discover many new high-redshift blazars.
GeV photons from high-redshift sources are expected to suffer
intergalactic absorption due to $e\pm$ pair production interactions
with the cosmic UV background radiation. 
The resulting features in the spectra of such blazars could therefore
be a key probe of the poorly-understood UV background radiation
and provide valuable insight into the cosmic dark ages \cite{oh01,gil09,sin09}.

The high-redshift evolution of the GLF is uncertain, as there are no
appropriate samples of gamma-ray blazars and X-ray AGNs above
$z=3$. Thus we constructed new GLFs by combining the U03 XLF
with evolutionary constraints from the Sloan Digital Sky Survey
(SDSS) quasar luminosity function data \cite{ric06}. The SDSS quasar
luminosity function is well constrained up to $z\sim5$.

Fig. \ref{fig.z_count} shows the expected cumulative redshift
distribution of blazars that are detectable by {\it Fermi} above 100 MeV.
Two sets of curves are plotted corresponding to the one- and five-year survey sensitivity.
Solid and dashed curves refer to the cases of whether or not
the SDSS quasar constraint is taken into account.
Note that intergalactic absorption effects are not included here,
as they are not expected to be important below 1 GeV \cite{oh01,gil09,sin09}.
The results imply that {\it Fermi} may detect some blazars up to $z\sim6$
during the five-year survey. Measurements of the spectra of such blazars
above 1 GeV through subsequent, deep observations should provide
interesting constraints on the high-z UV background.
More details will be presented in a forthcoming publication.

\section{Conclusions}
\label{section:conclusions}

We have compared our prediction of the EGRB spectrum \cite{IT09} with the
recent {\it Fermi} data, and made quantitative predictions for the next four years of the {\it Fermi} mission.
In Ref. \cite{IT09}, we constructed a model of the blazar gamma-ray
luminosity function (GLF), taking into account the blazar SED sequence
and the LDDE luminosity function inferred from X-ray observations of
AGNs. The GLF model parameters are constrained by fitting to the
observed flux and redshift distribution of the EGRET blazars. With this
model, we can predict the EGRB spectrum in a non-trivial way, rather than assuming simple power-law spectra.

The contribution from non-blazar AGNs to the EGRB is also considered,
using the nonthermal coronal electron model \cite{ITU08}. This model
gives a natural explanation for the observed cosmic MeV background,
and we examined whether the X-ray to GeV gamma-ray background radiation
can be accounted for by the two population model including blazars and
non-blazar AGNs.

Our predicted blazar EGRB spectrum matched very well with the latest
{\it Fermi} EGRB data. This indicates that blazars are the dominant
population of the EGRB and that other components such as dark matter
annihilation do not contribute to EGRB significantly. This also
indicates that the use of blazar SED sequence was valid. The predicted
spectrum including blazars and non-blazar AGNs is also in good
agreement with the observed data from X-ray to 100 GeV. Therefore,
blazars and non-blazar AGNs are likely the primary sources of
the cosmic X-ray and gamma-ray background radiation. Our model
predicts that it is possible to account for 80 \% of the observed EGRB flux
above 100 MeV with the sum of the blazar and non-blazar AGN components,
where blazars alone account for about 45 \%. The two components
are comparable at $\sim$100 MeV, with blazars and non-blazar AGNs
dominating at higher and lower energies, respectively.

From the viewpoint of the EGRB spectrum from MeV to GeV,
we argue that the dominant component contributing to the MeV background
is non-blazar AGNs that are also responsible also for the CXB,
rather than MeV blazars as proposed by Ref. \cite{aje09}.

We predicted the flux distribution of blazars by our model, and we
found that 750 and 1200 blazars in all sky should be detected by {\it
  Fermi}, assuming a sensitivity limit of $F_{\rm lim}=3\times
10^{-9}$ and $1\times 10^{-9}$ photons cm$^{-2}$ s$^{-1}$ above 100
MeV corresponding to one and five year survey sensitivity. This number
is significantly lower than the predictions by many of the previous
studies. About 4--50 of non-blazar AGNs are expected to be detected by
{\it Fermi} with 5-year survey.  {\it Fermi} should resolve almost all
of the EGRB flux from blazars into discrete blazars, with percentages
of $\gtrsim$98\% with 5-year survey.  On the other hand, less than
0.1\% of the EGRB flux from non-blazar AGNs can be resolved into
discrete sources.  Therefore, we have a clear prediction: {\it Fermi}
will resolve almost all of the EGRB flux into discrete sources at
photon energies $\gtrsim 1$ GeV where blazars are dominant, while a
significant fraction of the EGRB flux will remain unresolved in the
low energy band of $\lesssim$ 100 MeV where non-blazar AGNs have a
significant contribution.  This prediction can easily be tested, only
with the source counts and the EGRB estimates by {\it Fermi} data.

We also made predictions for future detections of high-redshift blazars.
Since the GLF is uncertain at $z \ge 3$ due to lack of direct observational data,
we constrained the redshift evolution of our original AGN XLF using SDSS quasar data.
With this assumption, we find that {\it Fermi} may find some blazars up to $z\sim6$ during the five year survey.
Such high-redshift GeV blazars will offer a new probe of the formation histories
of early stars and galaxies via GeV spectral attenuation signatures
caused by the high-redshift UV background radiation.

\bigskip 
\begin{acknowledgments}

 This work was supported by the Grant-in-Aid for the Global COE Program "The Next Generation of Physics, Spun from Universality and Emergence" and Scientific Research (19047003, 19047004, 19740099) from the Ministry of Education, Culture, Sports, Science and Technology (MEXT) of Japan. YI acknowledges support by the Research Fellowship of the Japan Society for the Promotion of Science (JSPS).

The authors wish to thank JACoW for their guidance in preparing
this template. Work supported by Department of Energy contract DE-AC03-76SF00515.
\end{acknowledgments}

\bigskip 

\end{document}